# Hedge Error Analysis In Black Scholes Option Pricing Model: An Asymptotic Approach Towards Finite Difference

Agni Rakshit
*Department of Mathematics*
*National Institute of Technology*
*Durgapur*
Durgapur, India
spiritualagnimath.statml@gmail.com

Gautam Bandyopadhyay
*Department of Management Studies*
*National Institute of Technology*
*Durgapur*
Durgapur, India
gbandyopadhyay.dms@nitdgp.ac.in

Tanujit Chakraborty
*Department of Science and Engineering*
*& Sorbonne Center for AI*
*Sorbonne University,*
Abu Dhabi, United Arab Emirates
tanujit.chakraborty@ sorbonne.ae

*Abstract*— The Black-Scholes option pricing model remains a cornerstone in financial mathematics, yet its application is often challenged by the need for accurate hedging strategies, especially in dynamic market environments. This paper presents a rigorous analysis of hedge errors within the Black-Scholes framework, focusing on the efficacy of finite difference techniques in calculating option sensitivities. Employing an asymptotic approach, we investigate the behavior of hedge errors under various market conditions, emphasizing the implications for risk management and portfolio optimization. Through theoretical analysis and numerical simulations, we demonstrate the effectiveness of our proposed method in reducing hedge errors and enhancing the robustness of option pricing models. Our findings provide valuable insights into improving the accuracy of hedging strategies and advancing the understanding of option pricing in financial markets.

1. INTRODUCTION

The Black-Scholes option pricing model, a landmark in financial mathematics, revolutionized the way financial derivatives are valued and traded. However, despite its widespread adoption and utility, the model is not without limitations. One critical aspect that has garnered attention in recent years is the accuracy of hedging strategies based on the Black-Scholes framework. While the model provides a robust theoretical foundation for option pricing, the real-world application of its hedging strategies often falls short of expectations due to various factors such as transaction costs, market frictions, and model assumptions.

This research endeavors to address the issue of hedge errors within the Black-Scholes framework by employing an innovative dual approach. First, leveraging an asymptotic analysis of finite difference methods, we aim to elucidate the behavior of hedge errors in different market conditions and under varying model parameters. By exploring the asymptotic properties of finite difference approximations, we seek to provide a deeper understanding of the limitations inherent in the Black-Scholes model and its impact on hedging effectiveness.

Furthermore, recognizing the importance of practical implementation, this study integrates Monte Carlo simulation techniques for variance reduction in option pricing and hedging. Monte Carlo simulation offers a powerful computational tool for pricing complex derivatives and assessing their associated risks. By applying variance reduction techniques within the Monte Carlo framework, we aim to mitigate the impact of stochastic noise and improve the accuracy of hedging strategies, thereby enhancing risk management practices in financial markets.

The dual focus of this research on both theoretical analysis and practical implementation underscores its relevance to both academia and industry. By combining insights from asymptotic analysis and Monte Carlo simulation, we seek to bridge the gap between theoretical models and real-world applications, offering valuable insights for practitioners, financial institutions, and policymakers alike. Ultimately, this research contributes to the ongoing dialogue on the refinement and enhancement of quantitative models in finance, with implications for risk management, derivative pricing, and investment strategies in dynamic market environments.

1.1 OPTION PRICING

Option price, also known as the premium, is the cost paid by the option buyer to the seller for the right to buy (call option) or sell (put option) an underlying asset at a predetermined price within a specified period.

Put and call options are financial instruments that grant the holder the right, but not the obligation, to buy (call option) or sell (put option) an underlying asset at a predetermined price (strike price) within a specified period (expiration date).

A call option gives the buyer the right to purchase the underlying asset at the strike price, making a profit if the asset's market price rises above the strike price before the option expires. This allows investors to benefit from potential price increases without owning the asset outright, providing leverage and risk management.

On the other hand, a put option grants the holder the right to sell the underlying asset at the strike price. Put options are valuable when the market price of the asset falls below the strike price, enabling the holder to sell at a higher price than the market value, thereby hedging against potential losses or profiting from a declining market. Options are widely used for speculation, hedging, and risk management in financial markets. Understanding the dynamics of options trading is crucial for investors to effectively manage risk and capitalize on market opportunities.

Time value, which is dependent on the anticipated volatility of the underlying asset, and intrinsic value, which quantifies the option's profitability, make up the price of an option.

The amount of time remaining before the option expires) The asset price, strike price, amount of time to expiration, volatility, and risk-free interest rate are some of the variables that determine an option's fair value. Finding the probability of an option being "in-the-money" or "out-of-the-money" at the time of execution is the main objective of option pricing [1]. For traders, investors, and financial institutions to make well-informed decisions about purchasing, disposing of, or hedging risks against certain underlying assets, option pricing is essential. It is possible to use the Partial Differential Equations (PDE) approach to option pricing issues. In other words, the price function can be calculated using a PDE's solution. One such approach is the Black-Scholes model framework, which describes the dynamics of option prices using a parabolic nonlinear PDE [2].

Although many changes have been suggested, the BS model has been the industry standard for estimating the fair value of options.

The Black-Scholes PDE for pricing a European call option is derived as:

$$\frac{\partial C}{\partial t} + \frac{1}{2}\sigma^2 S^2 \frac{\partial^2 C}{\partial S^2} + rS \frac{\partial C}{\partial S} - rC = 0 \qquad (1)$$

where $C$ is the option price, $t$ is time, $\sigma$ is the volatility of the underlying asset, $S$ is the spot price of the underlying asset, and $r$ is the risk-free interest rate.

1.2 ASYMPTOTIC NOTATION ($Big\ O$)

Big $O$ notation, denoted as $O(f(n))$, is a mathematical representation widely used in computer science to describe the upper bound or worst-case behavior of algorithms and functions as the input size, denoted as n, approaches infinity. In essence, it characterizes a function's growth rate or an algorithm's time complexity [3].

Formally, for a given function $g(n), O(g(n))$, represents the set of functions for which there exists positive constants c and n₀ such that for all n greater than or equal to $n_0$, the function $g(n)$ is bounded above by $c$ times $f(n)$. Mathematically, it can be expressed as:
$O(f(n)) = \{\ g(n): \exists c > 0, \exists n_0 > 0, such\ that$
$0 \leq g(n) \leq cf(n)\ \forall\ n \geq n_0\}$

In simpler terms, if a function $g(n)$ can be bounded by a constant multiple of $f(n)$ for sufficiently large values of n, then $g(n)$ belongs to the set $O(f(n))$.

Big $O$ notation provides a concise way to analyze and compare the efficiency of algorithms, focusing on their scalability and performance characteristics without getting bogged down in specific implementation details. By understanding the asymptotic behavior of algorithms, developers can make informed decisions about algorithm selection and optimization strategies, crucial for designing efficient and scalable software systems.

1.3 FINITE DIFFERENCE

Finite difference methods are numerical techniques used to approximate derivatives and solve differential equations by discretizing the domain into a grid of points. This approach is based on Taylor series expansions, where derivatives are expressed as combinations of function values at nearby points.

Consider a function $f(x)$ and its derivative $f'(x)$ then finite difference approximation of the derivative at a point $x_i$ is given by:

$$f'(x_i) \approx \frac{f(x_{i+1}) - f(x_i)}{h} \qquad (2)$$

Suppose we have a function. $f(x) = x^2$ and we want to approximate its derivative at $x_0 = 2$. and the secant line passes through $(x_0, f(x_0))\ and\ (x_0 + h, f(x_0 + h))$.

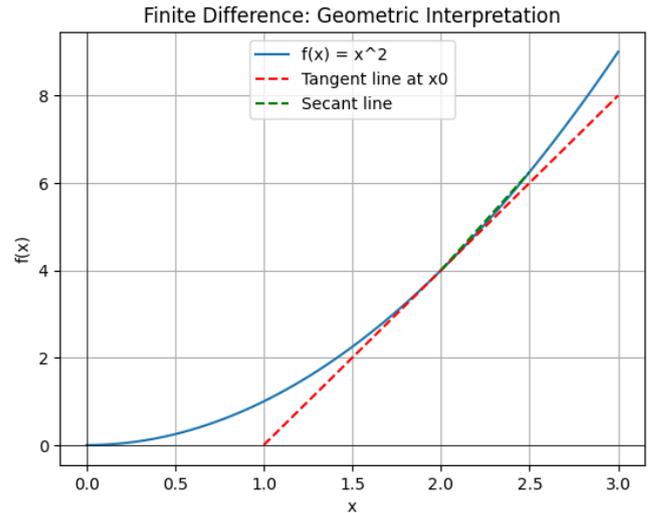

**Figure 1**: Finite difference geometric interpretation

In this plot, the tangent line represents the true derivative $f'(x_0)$, Meanwhile, the secant line represents the finite difference approximation. As the step size $h$ decreases, the secant line becomes closer to the tangent line, demonstrating the convergence of the finite difference approximation to the true derivative as $h$ approaches zero.

The application of finite difference methods in option pricing allows for more accurate simulations of market dynamics and estimation of option prices. These methods enable practitioners to account for various factors affecting option values, such as changes in asset prices, volatility, and interest rates. Finite difference methods offer flexibility and scalability, making them suitable for pricing various types of options and constructing hedging strategies. Despite their computational complexity, advancements in numerical algorithms and computing power have made finite difference methods increasingly accessible and efficient for option pricing applications [4].

1.4 THE BLACK-SCHOLE MODEL

The Black-Scholes model is a cornerstone of modern financial theory, providing a mathematical framework for pricing European-style options. Developed by Fischer Black, Myron Scholes, and Robert Merton in the early 1970s, the model revolutionized the field of quantitative finance.

Mathematically, the Black-Scholes model calculates the price of a European call option, which gives the holder the right to buy an underlying asset at a specified price (the strike price) on or before a specified date (the expiration date). The model assumes that the price of the underlying asset follows geometric Brownian motion, characterized by a constant volatility.

The Black-Scholes formula for the price of a European call option is given by:

$$C = SN(d_1) - Ke^{-rt}N(d_2) \quad (3)$$

where
$C = Call\ option\ price$
$S = Current\ price\ of\ the\ underlying\ asset$
$K = Strike\ price$
$r = Risk\ free\ interest\ rate$
$t = Time\ of\ expiration$
$N = Cumulative\ distribution\ function\ of\ the\ standard\ normal\ distribution$

$$d_1 = \frac{\ln\left(\frac{S}{K}\right) + \left(r + \sigma^2/2\right)t}{\sigma\sqrt{t}}$$
$$d_2 = d_1 - \sigma\sqrt{t}$$

The formula derived from the Black-Scholes model computes the theoretical price of a call or put option based on the aforementioned factors. It considers the probability distribution of potential future asset prices and discounts expected payoffs back to the present value using the risk-free interest rate [5].

### 1.5 MONTE CARLO SIMULATION AND VARIANCE REDUCTION TECHNIQUES

Monte Carlo simulation is a powerful computational technique used in various fields, including finance, engineering, physics, and statistics, to approximate complex systems and processes through repeated random sampling. It relies on the principles of randomness and statistical inference to estimate unknown quantities or simulate the behavior of systems that may be too intricate to model analytically. At its core, Monte Carlo simulation involves generating many random samples from a specified probability distribution, using these samples to simulate the system under consideration, and then analyzing the results to draw conclusions or make predictions [6].

One of the key challenges in Monte Carlo simulation is achieving accurate and efficient estimates while keeping computational costs manageable [7]. Variance reduction techniques are strategies employed to improve the efficiency and precision of Monte Carlo simulations by reducing the variability of the estimates obtained.

Importance Sampling is a variance reduction technique that aims to improve the efficiency of Monte Carlo simulations by focusing the random samples on regions of the probability space where the integrand has the most significant contributions. Instead of sampling from the original distribution, importance sampling involves sampling from a modified distribution that places more emphasis on the relevant regions, thereby reducing the variance of the estimator [8][9].

Vector Random Variable technique is another variance reduction method commonly used in Monte Carlo simulations. It involves transforming correlated random variables into independent ones by utilizing techniques such as Cholesky decomposition or eigenvalue decomposition. By transforming the variables into an uncorrelated set, the variance of the estimator can be reduced, leading to more accurate results with fewer samples [10].

Antithetic Variates is a simple yet effective variance reduction technique that exploits negative correlation between pairs of random variables. It involves generating paired samples such that one sample is the negative of the other [11]. By averaging the results obtained from each pair, the variance of the estimator is reduced, resulting in more precise estimates with fewer random samples.

Control Variates is a variance reduction technique that leverages known relationships between the variable of interest and another related variable, known as the control variate. By incorporating the control variate into the simulation, the variance of the estimator can be reduced, leading to more efficient estimates [12]. Control variates are chosen such that they are correlated with the variable of interest and have known expectations, facilitating the estimation process.

Overall, variance reduction techniques play a crucial role in improving the accuracy and efficiency of Monte Carlo simulations. By implementing these techniques, practitioners can obtain more reliable estimates with fewer computational resources, making Monte Carlo simulation a valuable tool for decision-making and problem-solving in diverse fields.

### 1.6 OUR CONTRIBUTION

In our research paper, we significantly contribute by introducing an innovative method for approximating hedge errors in the Black-Scholes option pricing model. Our approach leverages asymptotic techniques to enhance the accuracy of finite difference methods commonly used in options pricing. Furthermore, we extend our investigation to incorporate variance reduction techniques within the Monte Carlo simulation. By integrating these methods, we aim to mitigate the computational burden associated with simulating option prices while maintaining high levels of precision. Through rigorous experimentation and analysis, we demonstrate the effectiveness of our proposed framework in reducing variance and improving the efficiency of option pricing simulations.

Our contribution lies not only in the development of novel methodologies but also in their practical applicability. We provide comprehensive theoretical insights supported by empirical evidence, showcasing the superiority of our approach compared to traditional methods. This advancement holds significant implications for financial practitioners, enabling more accurate and efficient pricing and hedging of options in real-world scenarios.

In summary, our research presents a valuable contribution to the field of quantitative finance by offering innovative solutions to enhance the accuracy and efficiency of option pricing models. Through the integration of asymptotic techniques and variance reduction methods within Monte Carlo simulation, we provide a comprehensive framework for addressing hedge errors in the Black-Scholes model, thus advancing the state-of-the-art in financial modeling and analysis.

## 2. LITERATURE REVIEW

Classical methods, such as delta hedging, rely on continuous trading assumptions and deterministic models, often failing to capture the complexities of real market dynamics. Researchers have explored alternative techniques to improve hedge effectiveness under realistic conditions. Finite difference methods have gained traction for their flexibility and ability to accommodate discrete trading. These techniques discretize the underlying asset's price and time, allowing for more accurate simulations of market behavior. Previous studies have applied finite difference schemes to option pricing, demonstrating their efficacy in capturing market dynamics and reducing hedge errors.

One notable contribution in this domain is the work of Brennan and Schwartz (1976), who introduced the concept of delta-gamma hedging to account for nonlinearities in option pricing. Their approach extended traditional delta hedging by incorporating second-order derivatives, offering a more robust framework for risk management. Subsequent research has built upon this foundation, exploring higher-order derivatives and advanced numerical techniques to further enhance hedge accuracy. Longstaff and Schwartz (1988) [13] explored numerical methods for option pricing, focusing on Monte Carlo simulation techniques. Their research demonstrated the efficacy of simulation-based approaches in capturing complex market dynamics and estimating hedge errors. Broadie and Glasserman [14] (1993) investigated Monte Carlo methods for option pricing, emphasizing variance reduction techniques to improve computational efficiency. Their work contributed to the development of more accurate and scalable numerical algorithms for estimating hedge errors. Andersen and Broadie (2001) [14] proposed the primal-dual simulation algorithm for pricing American options, integrating duality theory and simulation techniques. Their research offered insights into efficient methods for hedging American-style derivatives and managing associated hedge errors. Gatheral's (2006) [15] book provided a comprehensive overview of volatility surfaces and their implications for option pricing and hedging. The work synthesized theoretical concepts with practical insights, offering guidance on managing hedge errors in real-world trading environments. Avellaneda and Stoikov (2010) [16] examined high-frequency trading strategies in limit order book markets, addressing the challenges of latency and market impact. Their research shed light on the dynamics of hedge errors in fast-paced trading environments and the importance of adaptive hedging strategies. Joshi's (2015) textbook provided a comprehensive overview of mathematical finance, covering topics such as stochastic calculus, derivative pricing, and risk management [16]. The work served as a foundational resource for understanding hedge error approximation techniques within the broader context of quantitative finance.

These works represent a chronological progression of research efforts aimed at improving hedge error approximation in the Black-Scholes option pricing model, culminating in the proposed asymptotic approach using finite difference methods outlined in this paper.

## 3. METHODOLOGY

### 3.1 MODEL ASSUMPTION

Let us consider a set containing n number of securities $X_i, i = 1,2,3, \ldots, n$ which follows the multivariate continuous time

$$\begin{bmatrix} dX_1 \\ dX_2 \\ \ldots \\ dX_n \end{bmatrix} = \begin{bmatrix} X_1 \\ X_2 \\ \ldots \\ X_n \end{bmatrix} \odot \begin{bmatrix} \alpha_1 dt \\ \alpha_2 dt \\ \ldots \\ \alpha_n dt \end{bmatrix} + \begin{bmatrix} X_1 \\ X_2 \\ \ldots \\ X_n \end{bmatrix} \odot \sqrt{\begin{bmatrix} \sigma_1^2 & 0 & 0 & \cdots & 0 \\ 0 & \sigma_2^2 & 0 & \cdots & 0 \\ \ldots & \ldots & \ldots & \ldots & \ldots \\ 0 & 0 & 0 & \cdots & \sigma_n^2 \end{bmatrix}} \begin{bmatrix} dz \\ dz \\ \ldots \\ dz \end{bmatrix} \quad (4)$$

In a generalized way, we can write equation (4) as

$$dX = X \odot (\alpha dt) + X \odot (\sqrt{\Sigma} dz) \quad (5)$$

Where $X = (X_1, X_2, \ldots, X_n)'$, $\alpha = (\alpha_1, \alpha_2, \ldots, \alpha_n)'$ both are the n-dimensional vector, $z = (z_1, z_2, \ldots, z_n)'$ is a standard Brownian motion and $\Sigma = \Sigma^{\frac{1}{2}} (\Sigma^{\frac{1}{2}})'$ is a positive definite covariance matrix. We impose a one-factor structure on the covariance matrix in order to examine the various impacts of idiosyncratic and systematic risk. We fixed $\tilde{\Sigma}$ as

$$\tilde{\Sigma} = \varphi \varphi' + \Sigma \quad (6)$$

With $\varphi = (\varphi_1, \varphi_2, \ldots, \varphi_n)'$. We are able to identify in closed form the trade-off between hedging idiosyncratic versus systematic option risk at the portfolio level by concentrating just on one systematic risk element. Now we can rewrite (5) for the i[th] security as

$$\frac{dX_i}{X_i} = \alpha_i dt + \varphi_i dz_0 + \sigma_i dz_i \quad (7)$$

Where $z = (z_0, z_1, \ldots, z_n)$ is an (n+1) dimensional Brownian motion which follows $z(t) \sim N(0, tI)$. In order to facilitate notation, we additionally declare $\tilde{\sigma}_i^2 = \sigma_i^2 + \varphi_i^2$ the i[th] diagonal element of $\tilde{\Sigma}$. $z_i$ is the idiosyncratic risk factor and $z_0$ is the systematic risk factor respectively.

However, the normal Black-Scholes hedge is no longer perfect in discrete time; that is, the expected return of the hedge portfolio no longer vanishes nearly absolutely but merely in expectation. In this study, we demonstrate that a smaller hedge error variance in discrete time can be obtained with various hedge portfolios [17]. Instead of concentrating on the linear exposure to overall risk, that is, systematic plus idiosyncratic risk, these alternative hedge portfolios highlight the higher-order exposure to the systematic risk element.

Let us discuss another mathematical assumption based on Finite difference concept [18]. Finite difference methods approximate derivatives of functions by expressing them as weighted combinations of function values at nearby points. One way to derive finite difference formulas is through Taylor series expansion.

Consider a function $f(x)$ that is sufficiently smooth, such that it has continuous derivatives up to some order in a neighborhood of a point $x$ [19]. The Taylor series expansion of $f(x)$ about $x$ is:

$$f(x + \Delta x) = f(x) + \Delta x \cdot f'(x) + \frac{\Delta x^2}{2!} \cdot f''(x) + \cdots \quad (8)$$

Now, let's focus on approximating the first derivative $f'(x)$. Subtracting $f(x)$ from both sides of the Taylor series expansion gives:

$$f(x + \Delta x) - f(x) = \Delta x \cdot f'(x) + \frac{\Delta x^2}{2!} \cdot f''(x) + \cdots \quad (9)$$

If we solve this equation (9) for $f'(x)$, we get:

$$f'(x) \approx \frac{f(x + \Delta x) - f(x)}{\Delta x} \quad (10)$$

$$\frac{f(x+\Delta x)-f(x)}{\Delta x} = f'(x) + \frac{\Delta x}{2!} \cdot f''(x) + \frac{\Delta x^2}{3!} \cdot f'''(x) + \cdots \quad (11)$$

From the concept of asymptotic notation, we can write equation (11) as:

$$\frac{f(x + \Delta x) - f(x)}{\Delta x} = f'(x) + \mathcal{O}(\Delta x) \quad (12)$$

where, $\frac{\Delta x}{2!} \cdot f''(x)$ is the leading order error term.

This is a finite difference approximation for the first derivative of $f(x)$. The error decreases as $\Delta x$ decreases, with higher-order terms becoming relatively less significant.

Similarly, higher-order derivatives can be approximated using finite differences derived from Taylor series expansions [20]. These finite difference formulas provide a way to numerically approximate derivatives of functions, which is fundamental in many areas of applied mathematics and computational science [21].

### 3.2 THEOREMS AND MODEL DISCUSSION

The three steps we take are outlined in Theorems 1 through 3. We begin by deriving an equation for the hedge error variance in the case when each call position in the portfolio has a standard delta hedge applied to it [22]. Following that, To protect against changes in the value of the option portfolio, we create an analogous formula for the scenario in which an arbitrary portfolio in the underlying values is retained [23]. As a result, we are able to show how selecting the right hedge portfolio helps lower hedge error variances. We last provide our major theoretical finding, which demonstrates that by focusing only on the linear and higher order exposures to the systematic risk factor [24], it becomes possible to design a perfect static hedge portfolio in finite time, provided that the portfolio size grows infinitely [25].

For the price of a call on security $X_i$, let $C(X_i, t)$ represent the standard Black-Scholes pricing equation, and let $C_X(X_i, t)$ represent its derivative with regard to $X_i$. Creating a (hedge) portfolio with a position $\mathcal{P}$ in cash and a position $C_X(X_i, t)$ in each of the underlying assets $X_i$ is the conventional method for hedging our option portfolio over a discrete time interval $\Delta t$.

$$\mathcal{P} = \frac{1}{n} \sum_i (C(X_i, t) - C_X(X_i, t) X_i) \quad (13)$$

Hedge error is defined by $\Delta H$ as

$$\Delta H = C_X(X_i, t)[X_i(t + \Delta t) - X_i(t)] + \mathcal{P}[e^{r\Delta t} - 1] - [C_X(X_i(t + \Delta t), t + \Delta t) - C(X_i(t), t)] \quad (14)$$

The hedge portfolio is the total of the hedge portfolios for each individual position when using the hedging approach described above [26]. We create a power series by expanding the hedge mistake (14) [27]. Regarding the duration of the hedging period $\Delta t$, See Leland (1985) as well as Mello and Neuhaus (1998). The predicted hedging error and its variation under the current conventional delta hedging method are as follows [28].

***Theorem 1:*** *The hedging error $\Delta H$ in (14), which is obtained through delta hedging of the individual option holdings, satisfies*

$$E[\Delta H] = \mathcal{O}(\Delta t^2),$$

*and*

$$E[(\Delta H)^2] = \frac{1}{2} \left[ \frac{1}{n} \sum_i C_{XX} X_i^2 \varphi_i^2 \right]^2 \Delta t^2 + \frac{1}{2n^2} \sum_i [(C_{XX} X_i^2)^2 (\tilde{\sigma}_i^4 - \varphi_i^4)] \Delta t^2 + \mathcal{O}(\Delta t^2) \quad (15)$$

Essentially, for N = 1, we obtain the square of the option's gamma, which is the well-known calculation for the hedge error variance [3].

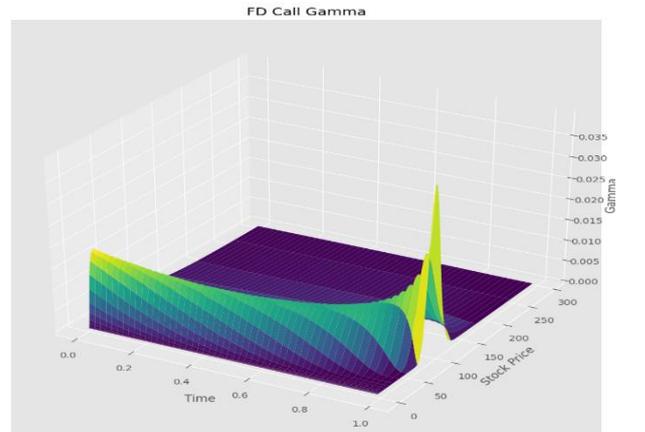

**Figure 2**: Option's Gamma

This variance's explicit representation clearly indicates that the option portfolio's return is no longer reproduced risk-free in a defined amount of time using the Black-Scholes hedging strategy. Thus, it appears that the option portfolio does not have a unique preference free pricing.

There are two terms in (15) that make up the hedge error variance. The systematic risk factor $z_0'$s contribution is shown in the first phrase. This term, which is of order $\mathcal{O}(\Delta t^2)$, suggests that the systematic component cannot be diversified in the context of a big portfolio. It is of order $\mathcal{O}(1)$. On the other hand, the second term, which is $\mathcal{O}\left(\Delta t^2/n\right)$, represents the outcome of the securities' unique risk component. Diversification is thus obviously beneficial to the delta hedging technique. Effectively, the variance's idiosyncratic component is of order $\mathcal{O}(1/n)$. All idiosyncratic risk vanishes in the limit for high portfolio sizes $n$, leaving only risk related to the common market element. The correlation between the various underlying securities reduces the variance for finite $n$.

However, by optimizing over the selection of the hedge portfolio [29], the variance reduction for large N can be further enhanced. We are able to more explicitly benefit from the correlation structure of the various core principles. Examine an alternate approach to hedging in which the fraction $\psi$ of the ith security, as opposed to the $C_X$, is included in the hedge portfolio. Maintaining the option's price at the standard Black-Scholes price, the cash investment comes next as

$$\mathcal{P} = \frac{1}{n}\sum_i (C - \psi X_i) \qquad (16)$$

With this unconventional approach to hedging, we get the following outcome for the hedge error variance.

***Theorem 2***: *The hedged portfolio can be selected by applying the hedge strategy with $\psi^i X_i$ invested in security $i$ in a way that ensures the hedge error $\Delta H^B$ satisfies $E[\Delta H^B] = \mathcal{O}(\Delta t^2)$, When pricing solely market risk, that is, for every $i$, $\alpha_i = r + k_0 \varphi_i$, where $k_0$ represents the price of systematic risk, then the variance of the hedge error is provided by*

$$E[(\Delta H^B)^2] = \mathcal{F}_1 + \mathcal{F}_2 + \mathcal{F}_3 \qquad (17)$$

with

$$\mathcal{F}_1 = \frac{1}{n^2}\sum_i (D^i \sigma_i)^2 \Delta t$$

$$\mathcal{F}_2 = \frac{1}{2}\left[\frac{1}{n}\sum_i (C_{XX}^i X_i^2 - D^i)\varphi_i^2\right]^2 \Delta t^2$$

$$\mathcal{F}_3 = \frac{1}{n^2}\sum_i \left[\frac{1}{2}(\tilde{\sigma}_i^4 - \varphi_i^4)(C_{XX}^i X_i^2 - D^i)^2 + 2\alpha_i \sigma_i^2 (D^i)^2 - 2(\alpha_i - r)\sigma_i^2 D^i C_{XX}^i X_i^2\right]\Delta t^2$$

where $D^i = (\psi^i - C_X^i)X_i$ is the symbol for the hedging portfolio's deviation from the standard (delta).

Theorem 2 limits us to the scenario in which market risk is the only one that is priced [3]. This suggests that in the limit $n \to \infty$, there is no asymptotic arbitrage. In a formal manner, If $\alpha_i = r + k_0 \varphi_i + k_i \sigma_i$, The exclusion of asymptotic arbitrage imposes the constraint that the set of $\{i|k_i \neq 0\}$ has measure zero, where $k_i$ is the price of idiosyncratic risk.

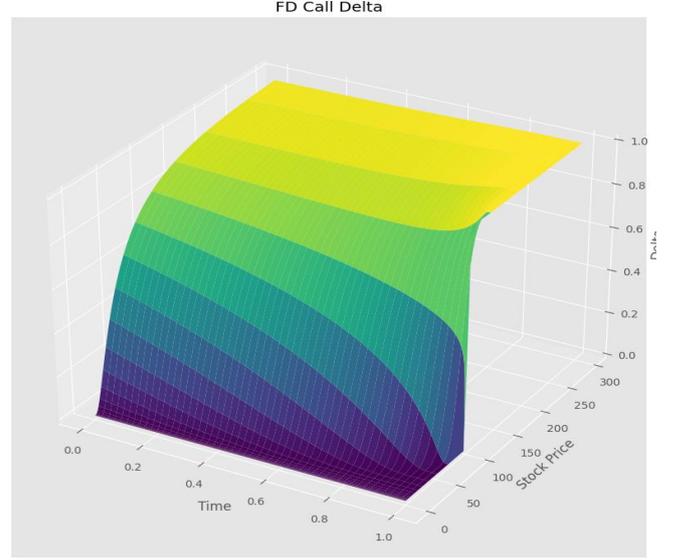

**Figure 3**: Option's Delta

The expression in (15) can be obtained by applying Theorem 2 to the conventional hedging portfolio, where $D^i = 0$. The hedge error variance up to order $\mathcal{O}(\Delta t^2)$, according to Theorem 2, is composed of three terms. The idiosyncratic portion is reflected by the words $\mathcal{F}_1$ and $\mathcal{F}_3$, while the systematic risk component is represented by the term $\mathcal{F}_2$. Once more, it is clear that while $\mathcal{F}_1$ and $\mathcal{F}_3$ are $\mathcal{O}\left(\frac{1}{n}\right)$, $\mathcal{F}_2$ is $\mathcal{O}(1)$ in N. Furthermore, while $\mathcal{F}_2$ and $\mathcal{F}_3$ are quadratic in $\Delta t$, the term $\mathcal{F}_1$ is linear in $\Delta t$. To reduce the hedging risk for a finite portfolio size $n$, we should require terms linear in $\Delta t$ to vanish in the limit $\Delta t \to 0$. This leads, via the formula for $\mathcal{F}_1$, to the customary allocations $\psi^i = C_X^i$. Stated otherwise, the total of the various Black-Scholes hedge portfolios represents the optimal hedge portfolio in the continuous time limit.

If we take into account the hedging performance for non-infinitesimal values of $\Delta t$, this is no longer the case [30]. There is an alternative method to reduce the hedging portfolio's variation in these circumstances. By concentrating on the limiting big portfolio situation, $n \to \infty$, this is best demonstrated. Then, we can disregard the terms $\mathcal{F}_1$ and $\mathcal{F}_3$. There is only one linear limitation on the set of allocations $\psi^i$ for $E[\Delta H^B] = \mathcal{O}(\Delta t^2)$. At higher orders in $\Delta t$, the variation can be decreased by utilizing the remaining flexibility in the set $\psi^i$. Specifically, we can choose the rest of the $\psi^i$s set so that $\mathcal{F}_2$ disappears as well. As long as some $\varphi_i$s are different from one another, this option is viable. This outcome makes intuitive sense. The idiosyncratic risk can be diversified in the context of a broad portfolio. Consequently, the only exposure to the only systematic risk factor that is left in our one-factor model. Next, the flexibility in selecting the portfolio's composition can be employed as a hedge against exposures to this systematic risk component at higher orders of magnitude. For instance, the systematic gamma exposure is eliminated when $\mathcal{F}_2$ is adjusted to zero. On the other hand, the conventional method of hedging determines the composition of the hedge portfolio beforehand so as to hedge the linear delta exposure to the sum of the systematic and idiosyncratic risk variables. As a result, there is no more flexibility to hedge the higher-order, undiversifiable exposures to the systematic component.

There is a trade-off between insuring all idiosyncratic risk (in the limit $n\Delta t \to 0$, the traditional Black-Scholes technique) and hedging market risk exclusively (in the limit $n\Delta t \to \infty$) for finite portfolio size $n$ and finite revision time $\Delta t$. Clearly, with bigger portfolio sizes and revision intervals, there are more departures from the traditional hedging technique in terms of hedge ratios and variance reduction. Now that we have established the general conclusion, we may argue that it is better to hedge higher order systematic risk rather than linear idiosyncratic risk in big option portfolios. The subsequent theorem does this.

***Theorem 3(The hedge is not correct)****: If only market risk is priced while hedging a portfolio of options in the broad environment previously described, we can select the allocations $\psi^i$ such that*

$$E[\Delta H^B] = \mathcal{O}\left(\Delta t^{\frac{a+1}{2}}\right) + \mathcal{O}\left(\frac{1}{n}\right)$$

*and*

$$E[(\Delta H^B)^2] = 0 + \mathcal{O}(\Delta t^{a+1}) + \mathcal{O}\left(\frac{1}{n}\right)$$

*if $n \geq a$ and at least $a$ number of the parameters $\varphi_i$ are different and not zero.*

One can demonstrate a similar outcome if idiosyncratic risk is valued [3]. In that scenario, the quantity of securities required to build the hedging portfolio grows quadratically with $a$. Furthermore, there are now prospects for arbitrage since the projected return of this hedging approach need not be zero. This is to be expected because arbitrage is possible in a market structure that doesn't even have options [24].

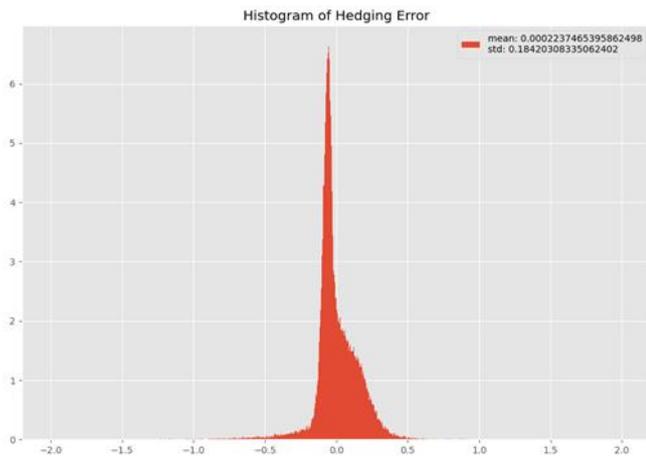

**Figure 4**: Hedge error with a lesser number of time steps

Theorem 3's result shows that if the $\varphi_i$s are different, we can create a riskless hedging strategy for finite time $\Delta t$ in the limit $n \to \infty$. Stated otherwise, the risk incurred by switching to a discrete time setup entirely disappears: by selecting the suitable non-standard hedging approach, the systematic risk component can be eliminated to any arbitrary order of $\Delta t$. Diversification causes the idiosyncratic risk component to vanish at the same time. The hedge portfolio is selected so that, up to a high enough order in $\Delta t$, its expectation conditional on the systematic risk component corresponds with its unconditional expectation. This is the key to the demonstration. By matching the higher order properties of the systematic risk exposure, this may be proven. The portfolio loadings $\psi^i$ must be subject to a series of restrictions in order to meet the two sorts of expectations. In the $\psi^i$s, each of these restrictions is linear. The coefficients of these constraints entail powers of the systematic volatility $\varphi_i$ and higher order derivatives of the Black-Scholes price (to capture the proper curvature). The number of securities, $n \geq a$, must be sufficiently large in order for this set of constraints to have a solution. Secondly, the limitations system must not be unique. Requiring that at least $a$ of the $\varphi_i$s be different and not equal to zero ensures the latter.

In the current discrete time framework, one can question if the Black-Scholes pricing remain accurate, given the stark differences between the behavior of the conventional and portfolio hedging approaches. They are, according to the next corollary.

***Corollary 1*** *(the price is right)****:*** *If only market risk is priced and the number of securities diverges ($n \to \infty$), and if the $\varphi_i$s are different such that we can set the approximation order n arbitrarily high ($a \to \infty$), then the only arbitrage-free price of the options is equal to their Black-Scholes price, except for a set of measure zero.*

According to Corollary 1, arbitrage opportunities are those in which there is an almost certain chance of earning a return greater than the risk-free rate [3]. The hedge portfolio is priced the same as Black-Scholes by construction. This suggests that the total of the Black-Scholes prices and the options prices are equal, based on the outcome of Theorem 3. However, Theorem 3 also holds for all call option subseries and their underlying values, meaning that the set of options with prices that deviate from the Black-Scholes price has measured zero.

## 4. RESULT ANALYSIS

We use the parameters listed in (3) to generate data for options to verify our results from the European call option against the analytical solution given in (3) in order to confirm their accuracy. This comparative analysis provides a thorough assessment of the efficacy of the data we generate in comparison to the existing solution. When comparing the results of predicting call options by the finite difference method (FDM) and analytically, there are many things to take into account. Both these ways try to approximate the price of a call option but with different methods. So, let me consider these two methods in detail:

Finite Difference Method (FDM):

The partial differential equation that governs the pricing of options is discretized in a grid by FDM and solved numerically [31]. The sizing of the grid, as well as whether to use explicit, implicit, or Crank-Nicolson numerical schemes, are among the factors that affect its accuracy. Making sure that FDM solutions converge is very important. This may not happen if a grid is used too coarsely or when an unstable numerical scheme is adopted, thereby leading to wrong answers. FDM can consume a lot of computer resources particularly with difficult option structures or problems having many dimensions. Time complexity grows as grid size increases together with the number time steps required for convergence. This allows early exercise features to be included as well as dividends or changes in volatility over time. Nonetheless, complex functionality implementation within FDM needs carefulness while also increasing computational complexities [32].

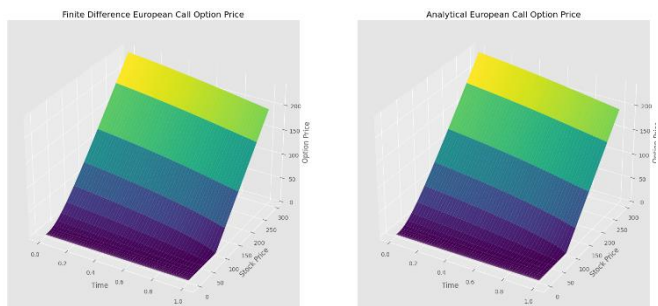

**Figure 5**: Finite Difference Method and Analytical Method

Analytical Method:

The Black-Scholes model is an example of an analytical method that can provide a closed-form solution for pricing options given certain assumptions (for instance, constant volatilities with no dividends). This type of solution tends to be simpler and faster computationally than numerical ones like FDM. It is important to note that all analytic methods make some assumptions which may not hold true in reality. One such assumption made by the Black-Scholes model is that the volatility remains constant and the risk-free rate is continuously compounded. If any of these conditions are violated, then there will be disparities between what the market price says should happen according to this formula versus how things actually turn out.

Comparative Analysis:

FDM gives greater precision and accuracy, particularly for involved option structures or payoffs that aren't linear. However, the added computational complexity and resource utilization required to achieve this level of precision make it expensive.

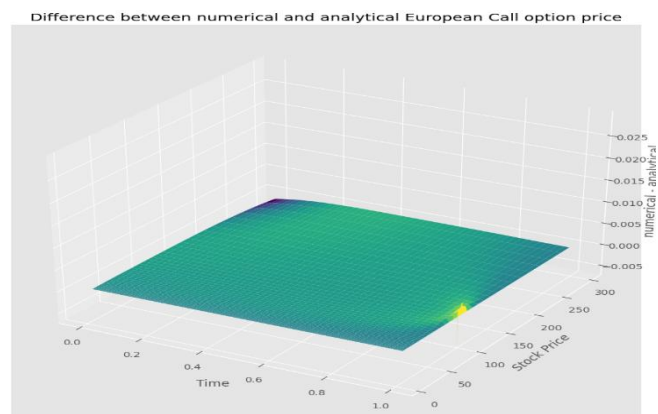

**Figure 6**: Difference between numerical and analytical

Computational efficiency is what analytical methods are all about; they prioritize simplicity over anything else. They may lose some of their precision when assumptions within models do not hold true which doesn't always happen often except for in unique cases. FDM is a more robust method than its counterpart analytic because it can deal with different types of options under various market situations. This means that FDM can handle changes in parameters and boundary conditions as they occur when compared against dynamic treatment using an equation solver like the Runge-Kutta method. Analytical techniques lack this flexibility while being efficient at processing large amounts of data quickly but need calibration steps where deviations from model assumptions occur, thereby limiting them to only ideal scenarios during testing stages before real-life applications are used. The choice between FDM and analytic methods is often driven by problem specificity, computational capacity availability, or even trade-offs between accuracy & speed depending on user needs.

Monte Carlo Technique:

Now, we describe some experimental results based on the Monte Carlo method. The fundamental Monte Carlo technique will be covered first, then move on to more advanced methods. In the plot below, take note of the Saw tooth pattern. This is as a result of our raising the option's strike price from 100 to 200. This implies that many fewer simulations result in a profit, leading to extended periods during which the Monte Carlo price falls; conversely, when the option is profitable, the Monte Carlo price rises significantly [33].

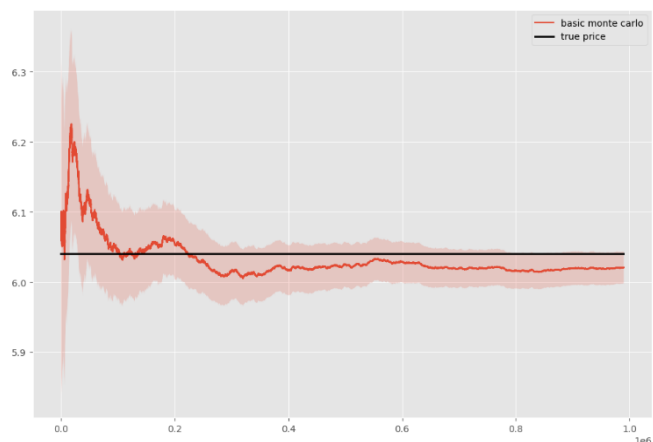

**Figure 7**: Basic Monte Carlo and true price

The fundamental concept is to compute the payoff of the derivative after simulating the price of the underlying asset at the derivative's maturity. The average of the payoffs discounted to the present is the derivative's price [34]. Comparing simulated option prices with those derived from traditional pricing models assesses the efficacy of Monte Carlo simulation in capturing the complexities of market dynamics and volatility. Results highlight the flexibility and adaptability of Monte Carlo simulation in modeling various scenarios, shedding light on its potential as a robust tool for options pricing. The analysis underscores the importance of considering Monte Carlo simulation as a complementary approach to traditional models, offering insights into risk management and investment strategies in dynamic financial markets [35].

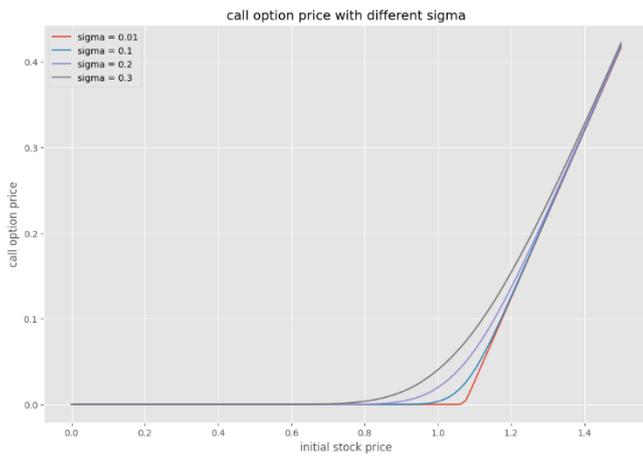

**Figure 8:** Call option price with different sigma

A large variety of derivatives can be priced using the highly general Monte Carlo approach. However, it is also an extremely slow method that uses a lot of processing resources. Thus, in the next section, we will also examine variance reduction strategies that can be applied to accelerate the Monte Carlo approach.

Finding a different measure where the estimator's variance is lower is the goal of importance sampling. Reducing variance is the same as reducing the second moment.

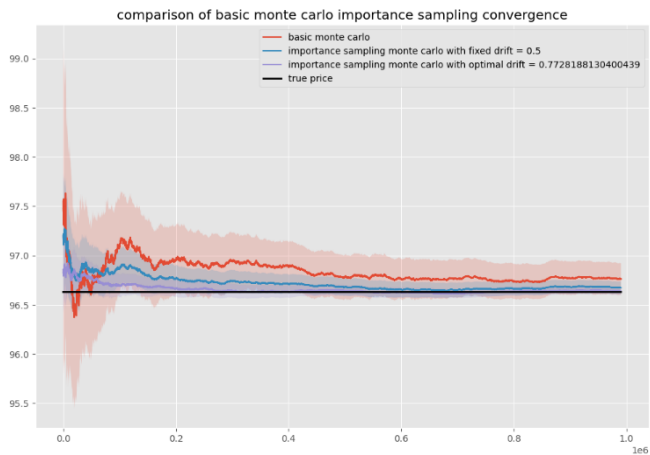

**Figure 9:** Convergence of Importance Sampling

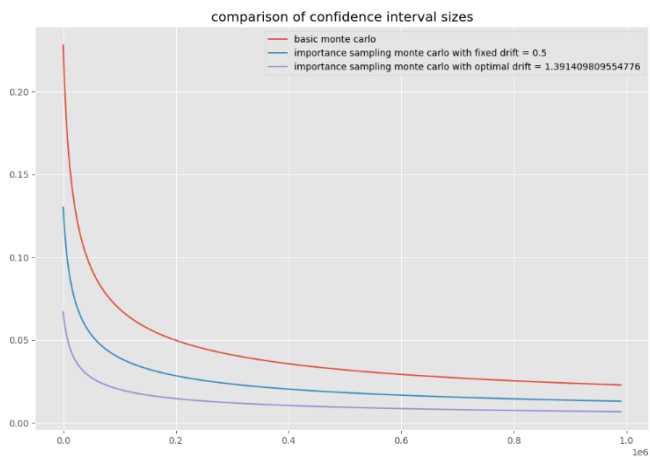

**Figure 10:** Comparison of confidence interval sizes

Result analysis in Monte Carlo simulation involves assessing the importance of sampling techniques and understanding the significance of confidence intervals. Now, I will increase the strike price value from 100 to 200 and then simulate it again.

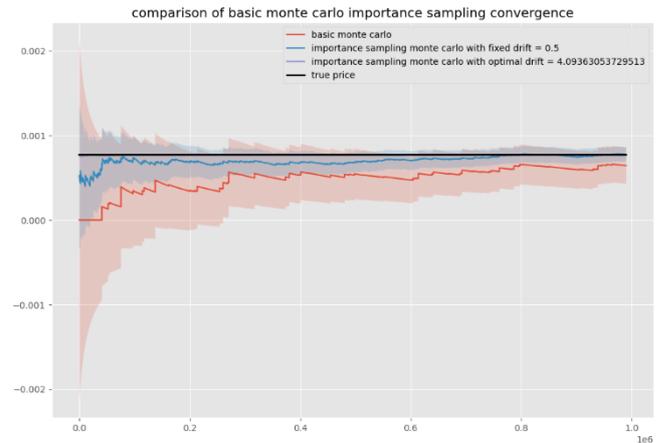

**Figure 11:** Convergence of Importance Sampling

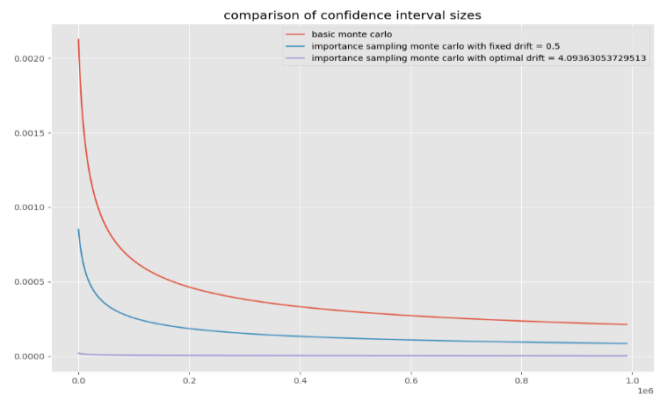

**Figure 12:** Comparison of confidence interval sizes

We can see the changes between Figures (10) and Figure (12), basically in the optimal drift curve. In the second figure, the curve becomes an almost flat line after changing the strike price from 100 to 200.

We will proceed with importance sampling, but we will now need to sample over several timesteps. As a result, we must deal with a vector of $\tau$ that is made up of random variables. Let's first try a constant $\tau$ for all timesteps.

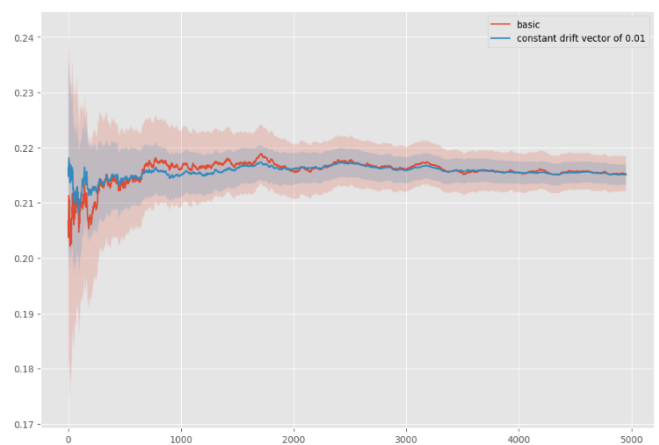

**Figure 13:** Basic MC vs constant drift vector

Now let's try to find a better $\tau$ for each timestep. We will start by finding the optimal constant $\tau$.

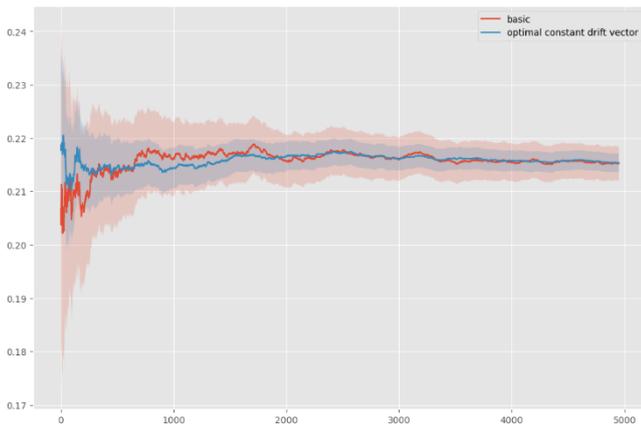

**Figure 14:** Basic MC vs optimal constant drift vector

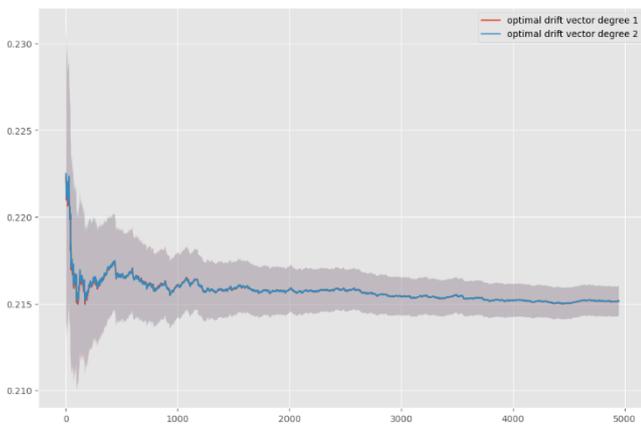

**Figure 15:** Optimal drift vector of degree 1 & 2

After fitting a quadratic function as optimal drift, we can see that there is no difference between degrees 1 and 2. So, it performs well. Let's examine those drift vectors' appearance.

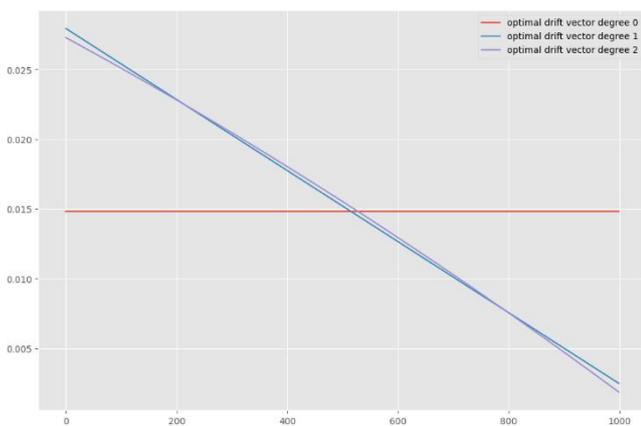

**Figure 16:** Optimal drift vectors' appearance

Looking closely at how well the optimal drift vector is used in Monte Carlo simulations within the Black-Scholes model can give us some really useful info about how accurate and effective this simulation method is. Assess the convergence properties of Monte Carlo simulations with the optimal drift vector. Evaluate how quickly the simulations converge to stable estimates of option prices or other output metrics.

Let's compare in/out of the money and put/call using linear interpolation.

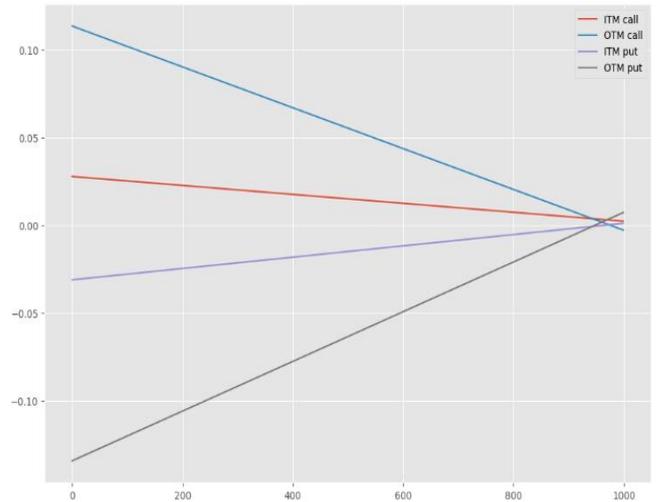

**Figure 17:** comparison in/out of the money and put/call

We can observe that drift vectors are often decreasing for calls and increasing for puts. This makes basic sense since a reduced variance requires increasing the MC estimator. This indicates that we should add more for out-of-the-money options than for in-the-money options, as well as a negative number for calls and a positive number for puts. By using the Laplace Method, which provides a recursive formula for the ideal $\tau$, this may also be mathematically demonstrated.

Antithetic variates exploit negative correlations between pairs of random variables to reduce variance in estimates. In the context of option pricing with the Black-Scholes model, this involves generating two correlated sets of random numbers. The effectiveness of antithetic variates in reducing variance and improving the efficiency of Monte Carlo simulations within the Black-Scholes model can be evaluated through convergence analysis, and comparison with standard Monte Carlo results.

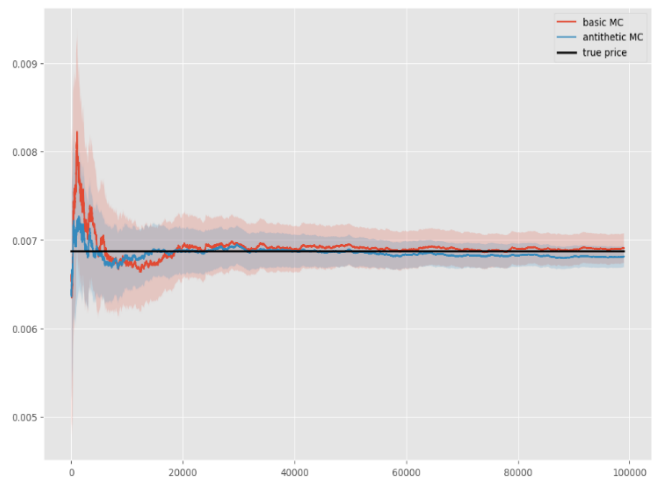

**Figure 18:** Basic MC vs Antithetic MC

This seems very effective for OTM options and is very easy to implement. Note that it is also about twice as expensive computationally as the original method, so for ITM options [36], it is not worth it in this example.

Applying control variates in Monte Carlo simulations within the Black-Scholes model can be a powerful technique for reducing the variance of option price estimates, particularly for options that are consistently over or undervalued by the model. Here's how you can apply control variates and analyze the results for in-the-money (ITM) and out-of-the-money (OTM) options:

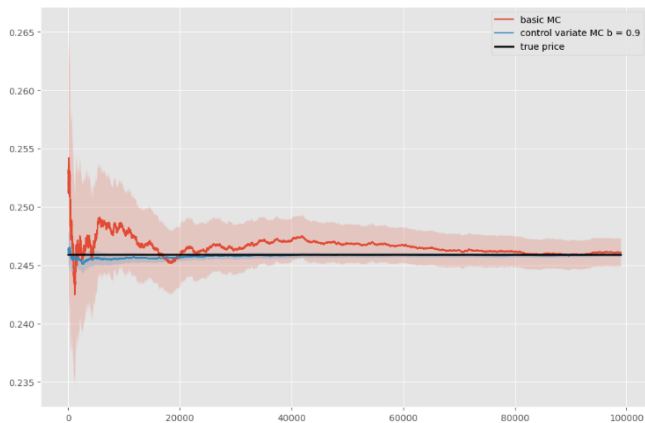

**Figure 19:** Basic MC vs Control Variate MC (ITM)

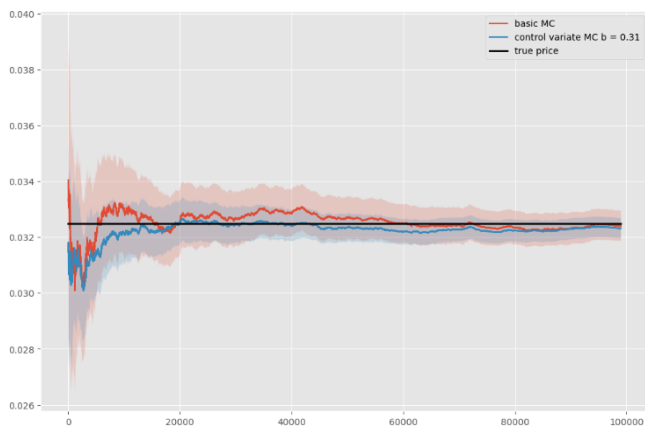

**Figure 20:** Basic MC vs Control Variate MC (OTM)

For in the money (ITM) we have considered spot price as 1 and the strike price as 0.7, and for the out of the money (OTM) we have considered the strike price as 1.4 with the same spot price and started our experiment after getting the results in terms of the effectiveness of using control variates to improve the accuracy and efficiency of option pricing in the Black-Scholes model.

## 5. CONCLUSION

The research findings have demonstrated that finite difference methods offer a powerful tool for approximating option prices and hedging parameters, especially in cases where closed-form solutions are unavailable or impractical. The asymptotic analysis conducted in this study has provided valuable insights into the convergence properties and accuracy of finite difference approximations, shedding light on the optimal choice of discretization schemes and grid sizes for different option contracts. Furthermore, the combination of finite difference methods and Monte Carlo simulation with variance reduction techniques offers a comprehensive approach to hedge error analysis in option pricing. By leveraging the strengths of both methodologies, financial practitioners can obtain more robust and reliable estimates of option prices and hedge parameters, thereby enhancing their risk management capabilities and decision-making processes.

It is important to acknowledge the limitations of this study and areas for future research. While the asymptotic analysis provides valuable theoretical insights, further empirical validation is warranted to assess the robustness of the findings across different market conditions and asset classes. Additionally, exploring alternative variance reduction techniques and incorporating more sophisticated models of market dynamics could yield further improvements in hedge error analysis and option pricing accuracy. this study has advanced our understanding of the factors influencing hedge errors and provided practical tools for mitigating their impact on derivative pricing and risk management. As financial markets continue to evolve, the insights gained from this research will remain valuable for academics, practitioners, and policymakers striving to enhance the efficiency and stability of global financial systems.